# Rounding of a first-order magnetic phase transition in Ga doped La$_{0.67}$Ca$_{0.33}$MnO$_3$

S. Rößler[1], U. K. Rößler[1], K. Nenkov[1], D. Eckert[1], S. M. Yusuf[2], K. Dörr[1], K.–H. Müller[1]

[1]Leibniz-Institut für Festkörper- und Werkstoffforschung IFW Dresden,
Postfach 270116, D-01171 Dresden, Germany.

[2]Solid State Physics Division, Bhabha Atomic Research Center,
Mumbai 400 085, India.

The effect of disorder on the critical properties of the ferromagnetic phase transition in colossal magnetoresistive manganite La$_{0.67}$Ca$_{0.33}$MnO$_3$ has been studied by substituting Ga for Mn. It is found that, upon 10 % Ga substitution, the peak in the specific heat at the Curie point $T_\mathrm{C}$ changes drastically and appears as a small anomaly. Static magnetization data analyzed in the asymptotic critical region using modified Arrott plots and the Kouvel-Fisher method give values for the critical exponents $\beta = 0.387(6)$, $\gamma = 1.362(2)$, and $\delta = 4.60(3)$. The results show that the first-order transition in La$_{0.67}$Ca$_{0.33}$MnO$_3$ becomes continuous by Ga substitution. The critical properties of the rounded transition in La$_{0.67}$Ca$_{0.33}$Mn$_{0.9}$Ga$_{0.1}$O$_3$ suggest that the magnetic subsystem in this mixed-valent perovskite is close to that of a conventional isotropic ferromagnet belonging to the Heisenberg universality class with short-range interactions. It is concluded that the first-order magnetic transition in pure La$_{0.67}$Ca$_{0.33}$MnO$_3$ is induced by fluctuations from a competing mode, which couples to the magnetic subsystem.

PACS numbers: 75.47.Lx, 64.60.Fr, 75.40.Cx, 75.50.Lk

**Introduction**

Colossal magnetoresistive (CMR) manganites of the type $R_{1-x}A_xMnO_3$, where R is a trivalent rare-earth ion and A is a divalent alkaline earth ion, has been a subject of intensive research in recent times due to their intricate magnetic and electrical properties.[1] For these compounds various electronic, magnetic and structural orders compete leading to very rich phase diagrams. Recently, the importance of multicritical points for their unconventional "colossal" behavior has been emphasized.[2-4]

For a certain range of compositions ($0.2 \leq x \leq 0.4$), the $R_{1-x}A_xMnO_3$ may show interdependent paramagnetic-ferromagnetic (pm-fm) and insulator-metal transitions. An explanation of this phenomenon has been given by a double-exchange model,[5] where electron hopping between Mn-$e_g$ levels align the Mn-$t_{2g}$ spins due to a strong intra-atomic Hund's coupling. For a quantitative explanation of CMR, additional mechanisms such as magneto-elastic couplings due to Jahn-Teller (JT) effects have been invoked.[6] Several experimental and theoretical works have demonstrated a close connection between JT distortions and localization of charge carriers or in other words, formation of polarons, in the insulating phase.[7,8] As a result, some of the theories associate CMR with a first-order pm-fm transition triggered by a discontinuous transition between polaronic and extended itinerant states of charge carriers.[9]

For the archetypical CMR-material $La_{0.67}Ca_{0.33}MnO_3$ clear evidence of a first-order pm-fm transition has been found, such as strong volume anomaly at the Curie point ($T_C$),[8] thermal hysteresis,[10] and negative slopes observed in magnetic isotherm plots $H/M$ vs. $M^2$.[11] The existence of a first-order pm-fm transition in the La$_{1-x}$Ca$_x$MnO$_3$ system is also corroborated by a tricritical point at the Curie-temperature $T_C$ for x=0.4.[12] On the other hand, a continuous phase transition has been observed in La$_{1-x}$Sr$_x$MnO$_3$ ($x = 0.125$, 0.3),[13,14] La$_{0.67}$Mg$_{0.33}$MnO3,[15] and Nd$_{0.6}$Pb$_{0.4}$MnO$_3$.[16] Values of the critical exponents characterizing the pm-fm transition in these compounds are close to those predicted for a 3D Heisenberg model with short-range interactions. This can be expected although the double exchange is driven by the motion of conduction electrons (Mn-e$_g$), the effective magnetic interaction near the transition is renormalized to isotropic short range interactions which couple the core-like Mn-t$_{2g}$-spins. Thus, the first-order and the continuous pm-fm transitions observed in different manganites indicate that the nature of the pm-fm transition seems to depend on the A-site doping.

The existence of a first-order pm-fm transition in La$_{0.67}$Ca$_{0.33}$MnO$_3$ is an intriguing problem in itself. It is known that thermally driven first-order transitions may be *rounded* by quenched disorder.[17-23] This means that a first-order transition in a pure system becomes continuous upon doping. Then, either a conventional critical behavior or novel types of continuous ordering transitions are observed in the disordered system. Thus, the existence of a first-order pm-fm transition in the La$_{1-x}$Ca$_x$MnO$_3$-system indicates that the inevitable mixed occupation of the A-site of the perovskite lattice with La and Ca does not affect the magnetic ordering-transition sufficiently to cause a continuous pm-fm transition. Interestingly, for a system with a continuous order parameter symmetry, as the magnetization in a Heisenberg-like magnet, there is a rigorous proof that, if the system undergoes a thermally driven first-order transition as a pure system, any amount of quenched disorder leads to a rounded continuous transition in spatial dimensions $d \leq 4$.[19,24]

In general, the coupling of the magnetic subsystem to other degrees of freedom may cause a fluctuation-induced first-order transition. Such a scenario has been analyzed in terms of phenomenological Landau-theory by using renormalization-group calculations for an isotropic magnetic system coupled to an Ising-like order parameter, which describes e.g. a charge order.[3] This analysis, together with the absence of first-order transitions in disordered isotropic magnets, suggests that the first-order pm-fm transition in $La_{0.67}Ca_{0.33}MnO_3$ is most probably close to a multicritical point where an additional fluctuating order parameter transforms the pm-fm transition into a first-order transition. Such a multicritical point with two competing orders could become a triple point where three lines of first-order-transitions meet and three different phases may coexist. Additional quenched disorder would round such a fluctuation-driven first-order transition in these systems. Then, continuous transitions will result with certain critical properties. For some theoretical models, where a fluctuation-driven first-order transition occurs, it has been shown that the critical properties of the transition rounded by quenched disorder will be the same as those of the corresponding pure system, i.e. a system without quenched disorder and decoupled from the fluctuating competing order parameter.[23] But, this need not be the case generally. In particular the transitions near multicritial points in the presence of quenched disorder are not well understood.[21] Therefore, it is interesting to study how strongly quenched disorder affects the pm-fm transition in $La_{0.67}Ca_{0.33}MnO_3$.

Here, we report how a direct substitution of non-magnetic ions in the magnetic sublattice affects the critical properties as compared to pure $La_{0.67}Ca_{0.33}MnO_3$. Such effects of a Mn-site substitution have not been studied so far. Studies on substituting a bigger ion like Ba or Sr for Ca in the A-site, i.e., $La_{0.67}(A_xCa_{1-x})_{0.33}MnO_3$ (where A = Ba or Sr )[25,11] have shown that a small amount of $A$ ion substitution result in a cross-over from a first-order transition to a continuous transition. The critical exponents obtained for the pm-fm transition in $La_{0.67}(Ba_{0.25}Ca_{0.75})_{0.33}MnO_3$ were found[25] in between those predicted for a 3D Heisenberg model and for mean field theory. However, such substitutions affect the lattice and modify the effective Mn-O-Mn bond angle, thereby changing the strength of the magnetic interactions. We introduce a substitutional ion Ga directly in the Mn-sublattice. $Ga^{3+}$ has a similar ionic radius as $Mn^{3+}$. Thus, this substitution avoids a static distortion of the lattice.[26] Also, since $Ga^{3+}$ has a filled shell configuration, it does not participate in the exchange interaction. Therefore, the observed change of the pm-fm transition is related only to the random impurities, viz. site-dilution, in the magnetic subsystem and suppression of dynamic JT distortion since $Ga^{3+}$ is not a Jahn-Teller ion. We show that the pm-fm transition becomes continuous in $La_{0.67}Ca_{0.33}Mn_{0.9}Ga_{0.1}O_3,$ upon 10 % Ga substitution for Mn in $La_{0.67}Ca_{0.33}MnO_3$. Three critical exponents associated with the static magnetization behaviour at the transition are evaluated independently. The values of critical exponents are close to those expected for 3D Heisenberg model with short-range interactions.

**Experimental**

Polycrystalline $La_{0.67}Ca_{0.33}Mn_{1-x}Ga_xO_3$ ($x = 0$ and $x = 0.10$) samples were prepared by a conventional ceramic method. Stoichiometric amounts of $La_2O_3$, $CaCO_3$, $MnC_2O_4·2H_2O$, and $Ga_2O_3$ were ground well, and the homogeneous mixture was heated at 900°C for 24 h, cooled to room temperature, reground, and heated at 1250°C for 24 h. The black powder thus obtained is pelletized and sintered at 1500°C for 12 h. The phase purity was investigated by x-ray diffraction. Both the compounds crystallize in orthorhombic structure with the space group *Pbnm*. Resistance measurements done by standard four-probe method showed insulator-metal transitions with resistivity peak-temperatures of about 253 K and 100 K for $x=0$ and $x=0.10$ compositions, respectively. The specific heat measurements were carried out in a model 6000 physical property measurement system (PPMS, Quantum Design) down to 2 K. Extensive magnetization measurements $M(T,H)$ were performed on $x=0$ and $x=0.10$ samples in external static magnetic fields H up to 48 kOe in the temperature range encompassing the respective critical regions (230-260 K for $x=0$ and 110-130 K for $x=0.10$) near the paramagnetic-ferromagnetic phase transition using a SQUID magnetometer. The data were collected in temperature steps of 0.5 K.

**Results and Discussion**

The temperature dependence of magnetization measured in a field of 0.3 T of $La_{0.67}Ca_{0.33}Mn_{1-x}Ga_xO_3$ ( $x=0$ and 0.10) is shown in Fig. 1. The sharp transition for $x=0$ indicates a first order transition. The Curie temperature, defined as a point where $dM/dT$ has a minimum, was found to be 240 and 120 K for $x=0$ and $x=0.10$, respectively. In Fig. 2, specific heat as a function of temperature is plotted. A sharp peak observed at $T=236$ K in the specific heat ($C_p$) measurements of $x=0$ sample is also consistent with the first-order transition. For $x=0.10$, on the other hand, no visible peak or cusp was observed in the specific heat curve. However, a small anomaly in specific heat around the Curie point can be seen when $C_p/T$ vs. $T$ is plotted as shown in the inset of Fig. 2. This behavior is similar to that observed in $Nd_{0.6}Pb_{0.4}MnO_3$, which showed a continuous transition.[27] The ferromagnetic transition also is broadened (Fig 1) suggesting a smearing of transition due to Ga impurities in the Mn-sublattice. Since the specific heat behavior did not indicate a first order transition for $x = 0.10$ sample, we tried to analyze the static magnetization data assuming a continuous phase transition for this compound.

A continuous phase transition near the critical temperature $T_C$, according to the scaling hypothesis, shows a power law dependence of spontaneous magnetization $M_S(T)$, and inverse initial susceptibility $\chi_0^{-1}(T)$ on the reduced temperature $\varepsilon = (T - T_C)/T_C$ with a set of interdependent critical exponents $\beta$, $\gamma$, and $\delta$ etc.[28] as given below.

$$M_S(T) = M_0(-\varepsilon)^\beta, \qquad \varepsilon < 0, \tag{1}$$

$$\chi_0^{-1}(T) = (h_0/M_0)\varepsilon^\gamma, \qquad \varepsilon > 0, \tag{2}$$

At $T_C$ exponent $\delta$ relates $M$ and $H$ by

$$M = D H^{1/\delta}, \qquad \varepsilon = 0, \tag{3}$$

Here, $M_0$, $h_0/M_0$ and $D$ are the critical amplitudes. Further, the scaling hypothesis predicts that $M(H, \varepsilon)$ is a universal function of $T$ and $H$,

$$M(H, \varepsilon) = \varepsilon^\beta f_\pm(H/\varepsilon^{\beta+\gamma}), \tag{4}$$

where $f_+$ for $T > T_C$ and $f_-$ for $T < T_C$, respectively, are regular functions. Eq. (4) implies that $M/\varepsilon^\beta$ as a function of $H/\varepsilon^{\beta+\gamma}$ falls on two universal curves, one for temperatures above $T_C$ and the other for temperatures below $T_C$.

In the case of the $x = 0$ sample, the field dependence of magnetization measured in the vicinity of the Curie point showed anomalies in the slopes similar to a metamagnetic transition. Such change in slope, according to the criterion given by Banerjee,[29] is used to distinguish a first-order transition from continuous ones by purely magnetic methods. This method is assumed by Mira et al.[11] to distinguish the nature of pm-fm phase transition in $La_{0.67}(Sr_xCa_{1-x})_{0.33}MnO_3$. They found a negative slope of isotherm plots of $H/M$ vs. $M^2$ for $La_{0.67}Ca_{0.33}MnO_3$ characterizing a first-order transition. We also found similar results. On the other hand, no such anomalies were found in $La_{0.67}Ca_{0.33}Mn_{0.9}Ga_{0.1}O_3$. Hence we proceeded with the scaling analysis.

In order to extract the spontaneous magnetization and susceptibility for $x = 0.10$ from the $M-H$ isotherms, we constructed the Arrott plot $M^2$ vs. $H/M$ in Fig. 3 after correcting the external magnetic field for the demagnetization effect. Such curves should give a series of straight lines for different temperatures and the line at $T = T_C$ should pass through the origin, according to the mean-field theory. In the present case, the curves were found to be non-linear suggesting that the mean-field theory is not valid. Therefore we analyzed the data using a modified Arrott plot,[30] in which $M/\varepsilon^{\beta'}$ is plotted versus $(H/M)^{1/\gamma'}$ as shown in Fig 4. Different values of $\beta'$ and $\gamma'$ were taken as trial values for the construction of the modified Arrott plot. If the system is close to a tricritical point as in the case of $La_{0.6}Ca_{0.4}MnO_3$, then mean-field exponents for tricritical points $\beta = 0.25$ and $\gamma = 1$ [12] are expected. In the extreme disorder limit, the exponents should reach the Fisher-renormalized tricritical exponents ($\beta = 0.5$ and $\gamma = 2$).[31,32] By checking these different possibilities, we find that the best description with nearly linear behavior for fields $1 < H < 50$ kOe is obtained in the modified Arrott plots for $\beta' = 0.365$ and $\gamma' = 1.336$. From a linear extrapolation from fields above 1 kOe to the intercepts with the axes $M^{1/\beta'}$ and $(H/M)^{1/\gamma'}$, values of spontaneous magnetization $M_S(T,0)$ and inverse susceptibility $\chi_0^{-1}(T,0)$, respectively, can be extracted. The isothermal line that passes through the origin is the critical isotherm at $T = T_C$. These values, $M_S(T,0)$ and $\chi_0^{-1}(T,0)$, are then plotted as functions of temperature. The power law fits according to Eqs. (1) and (2) to $M_S(T,0)$ and $\chi_0^{-1}(T,0)$, respectively, give the values of $\beta$ (eq. (1)) and $\gamma$ (eq. (2)). These new values of $\beta$ and $\gamma$ are then used to construct new modified Arrott plots. The iteration was continued until stable values of $\beta$, $\gamma$, and $T_C$ were obtained. In Fig. 5, $M_S(T,0)$ and $\chi_0^{-1}(T,0)$ versus temperature are plotted. The continuous curves show the power law fits according to eqs. (1) and (2) to $M_S(T,0)$ and $\chi_0^{-1}(T,0)$, respectively. This gives the values of $\beta = 0.380(2)$ with $T_C = 116.19(2)$ K (eq. (1)) and $\gamma = 1.365(8)$ with $T_C = 116.11(2)$ K (eq. (2)). The exponents obtained in this way are close to those expected for a short-range Heisenberg model ($\beta = 0.3680(3)$, $\gamma = 1.3960(9)$).[33]

In order to obtain more precise values of critical exponents, the Kouvel-Fisher (KF) method[34] has been used. In this method, plots of $M_S(dM_S/dT)^{-1}$ vs. $T$ and $\chi_0^{-1}(d\chi_0^{-1}/dT)^{-1}$ vs. $T$ (Fig. 6) should yield straight lines with slopes $1/\beta$ and $1/\gamma$, respectively. When extrapolated to the ordinate equal to zero, these straight lines should give intercepts on their $T$ axes equal to the Curie temperature. The straight lines obtained from a least-square fit to the data give the values of $\beta = 0.387(6)$, $T_C = 116.05(6)$ K and $\gamma = 1.362(2)$, $T_C = 115.88(6)$ K, respectively.

To obtain the value of $\delta$, the critical isotherm $M_S(116\,\text{K}, H)$ vs. $H$ on a log-log scale has been plotted in Fig 7. According to eq. (3), this should be a straight line with slope $1/\delta$. From the linear fit we obtained $\delta = 4.60(3)$, which compares well with the value $\delta = 4.783(3)$ expected for a Heisenberg-like ferromagnet.[33]

The scaling equation also predicts $\delta = 1 + (\gamma/\beta)$.[35] From the values of $\beta$ and $\gamma$ obtained from Fig. 5, the calculated value of $\delta$ according to this equation is 4.59 and that from KF method is 4.52. The value of $\delta$ independently obtained from magnetization isotherm (Fig 7) is in agreement with the scaling hypothesis within the experimental errors.

A more stringent test for scaling is to plot $M/\varepsilon^{\beta}$ vs. $H/\varepsilon^{\beta+\gamma}$ and to see whether the data obey the scaling equation of state (Eq. (4)). If Eq. (4) holds, all of the data should fall on one of the two curves. By taking the values of $\beta$ and $\gamma$ obtained from KF method and $T_C = 116$ K, the scaled data are plotted in Fig. 8 on a log scale. It can be clearly seen that all the data fall on two curves, one for $T > T_C$, and the other one is for $T < T_C$. Thus, our result suggests that doping in the magnetic sub-lattice of $La_{0.67}Ca_{0.33}MnO_3$ induces a change from a first-order to a continuous phase transition.

In the case of materials that show a continuous phase transition in the pure limit, the effect of quenched disorder is ruled by the Harris criterion,[36] which can be justified by renormalization group calculations.[37,38] The theory predicts that quenched random impurities do not alter the static critical exponents at a continuous transition, if the specific heat exponent $\alpha$ of the pure system is negative (α < 0 as in Heisenberg-like ferromagnets), while in the opposite case, a crossover from a pure to a random fixed point with new critical exponents occurs. These predictions have been confirmed experimentally.[32,39,40] This mechanism is closely related to the critical properties, which arise at a rounded first-order transition.[23] The critical properties found here for $La_{0.67}Ca_{0.33}Mn_{0.9}Ga_{0.1}O_3$ can be understood as those of a conventional isotropic ferromagnetic system where quenched disorder is not relevant.

There are few studies on pm-fm phase-transitions, where the pure compound shows a first-order transition. In the perovskitic manganites, the only compound that has clearly been identified to show a first-order pm-fm transition is $La_{2/3}Ca_{1/3}MnO_3$.

Observation of this first-order transition in $La_{2/3}Ca_{1/3}MnO_3$ and a continuous one in $La_{2/3}Sr_{1/3}MnO_3$ led some of the authors to distinguish $La_{2/3}Ca_{1/3}MnO_3$ from other manganites as a low temperature CMR material. It has been argued[41] that a phase-coexistence is a necessary condition for the CMR phenomenon. However, recently it has been shown that another low temperature material, metallic $Nd_{0.6}Pb_{0.4}MnO_3$ with $T_C = 156$ K and a sizeable CMR, has a continuous phase transition.[16] Some of the insulating manganites such as $La_{0.67}Mg_{0.33}MnO_3$ [15] and $La_{0.875}Sr_{0.125}MnO_3$ [13] also show a continuous pm-fm transition with critical properties resembling those predicted for Heisenberg ferromagnets with short-range interactions. This comparison shows that the pm-fm transition in $La_{2/3}Ca_{1/3}MnO_3$ is different from those of other materials in this class. As discussed above, the intrinsic disorder, if it is important at all, should allow only a continuous pm-fm transition in these systems because the magnetic order-parameter, i.e. a magnetization vector with three components, has continuous symmetry.[19,21] Our result of a rounded transition with conventional critical indices in $La_{2/3}Ca_{1/3}MnO_3$ with Mn-site doping demonstrates this true effect of quenched disorder. Hence, the existence of a first-order pm-fm-transition in pure $La_{2/3}Ca_{1/3}MnO_3$ must be related to another competing ordering mechanism leading to a fluctuation-induced first-order behaviour. Thus, our result can be understood as the rounding of the first-order transition near a multicritical (triple) point. In fact, recent powder neutron diffraction data for $La_{0.67}Ca_{0.33}MnO_3$ display the coexistence of paramagnetic, ferromagnetic and antiferromagnetic/charge-ordered domains in a broad temperature range around the transition.[42] Alonso et al.[43] have investigated a disordered model for $La_{3/2}Ca_{1/2}Mn_{1-x}Ga_xO_3$ and predict a quantum critical point for $x = 0.1$ to $0.2$ associated to the localization of the electronic states of the conduction band. This scenario is similar to a multicritical phase-diagram where quenched disorder drives down the transition temperature to zero and a disorder-line phase diagram occurs.[21] On the other hand, also a decoupled tetracritical phase-diagram may result with coexistence of different magnetic phases. This may arise because, for stronger disorder in the range above $x > 0.1$, another type of magnetic order should occur which may correspond to a random fixed point in the language of the renormalization group.[37,38] In an intermediate range of disorder, both magnetic phases may co-exist. Such a phase diagram would show a magnetic transition at higher finite temperatures either into the ferromagnetic or the random phase, while a second critical line appears at lower temperatures, when entering the co-existence region. In earlier experiments on $La_{3/2}Ca_{1/2}Mn_{1-x}Ga_xO_3$ randomly canted ferromagnetism for $x = 0.1$ and features of a cluster-glass with a remaining ferromagnetically ordered component for $x = 0.25$ were found.[26] This indicates rather a decoupled tetracritical phase diagram where ferromagnetic order coexists with some random type of magnetic structure, which is enforced by the quenched disorder.

**Conclusions**

Effect of disorder on the critical properties of optimally doped $La_{2/3}Ca_{1/3}MnO_3$ has been studied by specific heat and static magnetic measurements in the critical region. It is shown that the substitution of non-magnetic impurities softens the ferromagnetic transition and suppresses the phase coexistence. This disorder effect provides an important input in understanding ferromagnetism, metallicity, and the role of couplings to the lattice in the intensely studied, yet not well understood, manganite $La_{2/3}Ca_{1/3}MnO_3$. The effect of the Ga-doping causes the expected rounding of the first-order transition and reveals the usual critical behavior at a ferromagnetic ordering. The estimated critical exponents indicate that the magnetic subsystem of the disordered compound likely belongs to the universality class of conventional three-dimensional Heisenberg-like ferromagnets with short-range interactions. This would mean that, from a phenomenological point of view, the magnetic subsystem of $La_{2/3}Ca_{1/3}MnO_3$ is essentially that of a conventional isotropic ferromagnet. However, there must be a strong coupling to other modes which causes the fluctuation-driven first-order magnetic transition in the pure compound.

**Acknowledgments**

SR thanks the Alexander von Humboldt foundation for financial support. UKR and SR are grateful to R. Narayanan and N. Shannon for discussions.

**Figure Captions**

1. Temperature dependence of magnetization for $La_{0.67}Ca_{0.33}Mn_{1-x}Ga_xO_3$ ($x = 0$ and 0.10) measured in a dc field of 3kOe.

2. Specific heat as a function of temperature of $La_{0.67}Ca_{0.33}Mn_{1-x}Ga_xO_3$ ($x = 0$, and 0.10). The arrow shows the Curie point of $x = 0.10$. In the inset, $C_p/T$ vs. $T$ is shown for $x = 0.10$ sample around the Curie temperature.

3. Isotherms $M^2$ vs. $H/M$ of $x = 0.10$ sample at different temperatures close to the Curie temperature ($T_C = 116$ K).

4. Modified Arrott plot isotherms $M^{1/\beta'}$ vs. $(H/M)^{1/\gamma'}$ for $x = 0.10$ sample, with trial values $\beta' = 0.365$ and $\gamma' = 1.336$, $T = T_C = 116$ K is the value obtained in this study.

5. Temperature variation of the spontaneous magnetization $M_S$ and the inverse initial susceptibility $\chi_0^{-1}$ along with fits obtained for power laws explained in the text for $x = 0.10$ sample.

6. Kouvel-Fisher plot for the spontaneous magnetization $M_S$ and the inverse initial susceptibility $\chi_0^{-1}$ for $x = 0.10$ sample.

7. $M$ vs. $H$ on a log-log scale at 116 K, i.e. $T = T_C$, for $x = 0.10$ sample. The straight line is the linear fit following eq. (3).

8. Normalized isotherms of $x = 0.10$ sample below and above Curie temperature ($T_C = 116$ K) on a log-log scale using $\beta$ and $\gamma$ determined as explained in the text. $|\varepsilon| = |T - T_C|/T_C$.

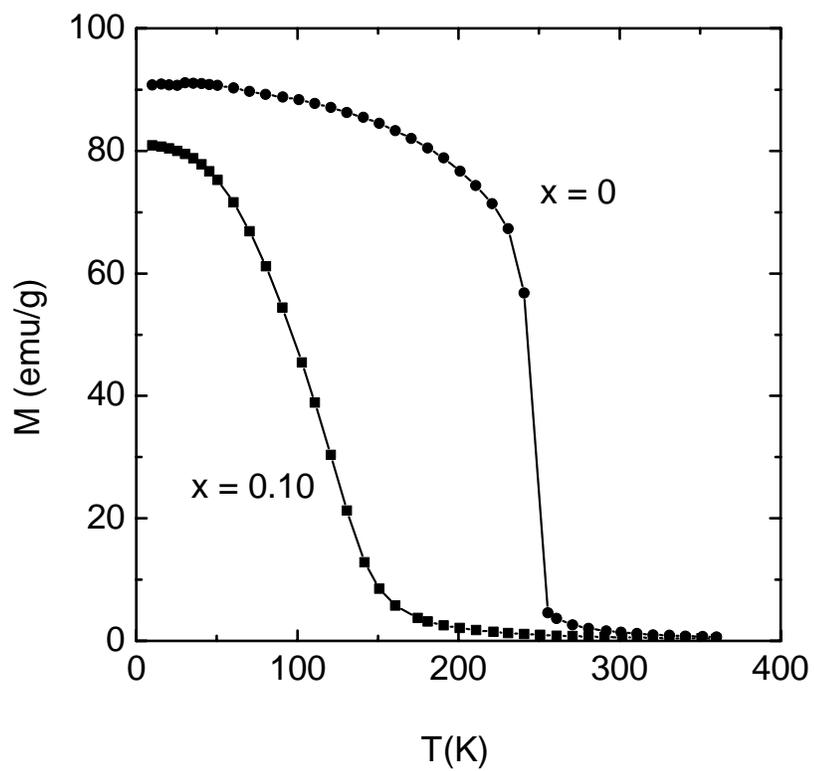

**Fig. 1**

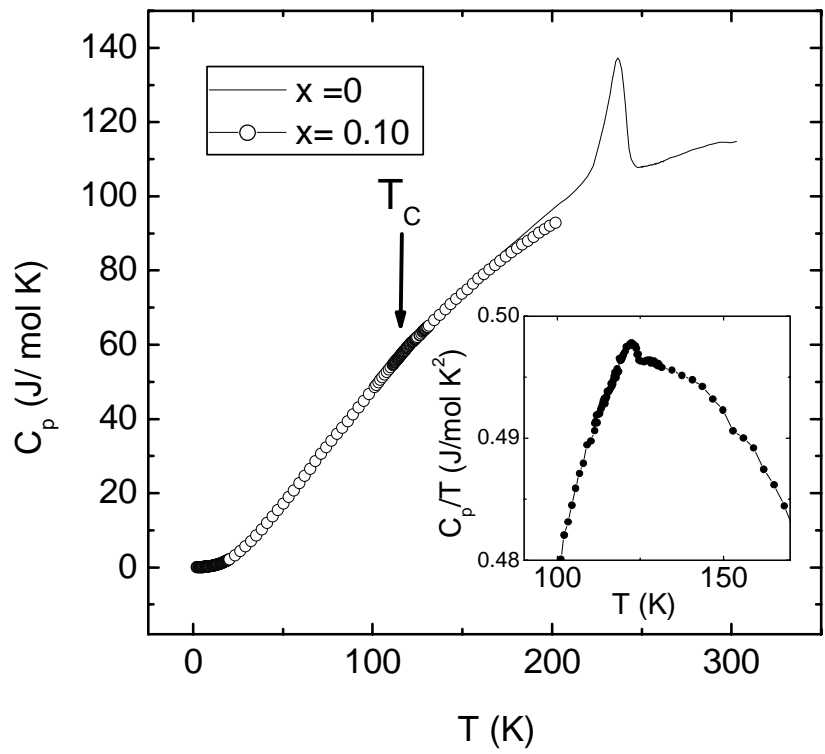

**Fig. 2**

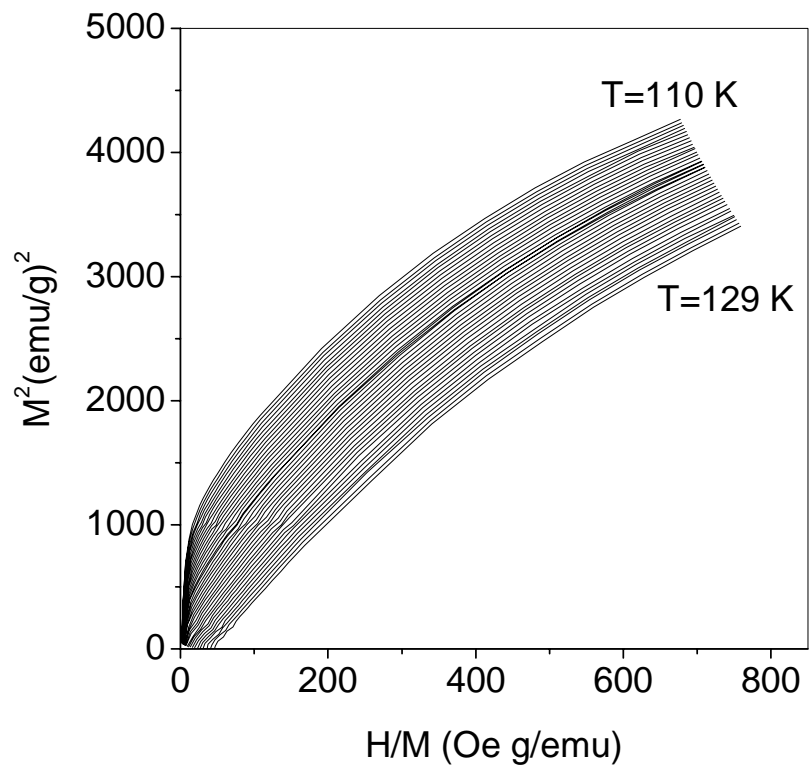

**Fig. 3**

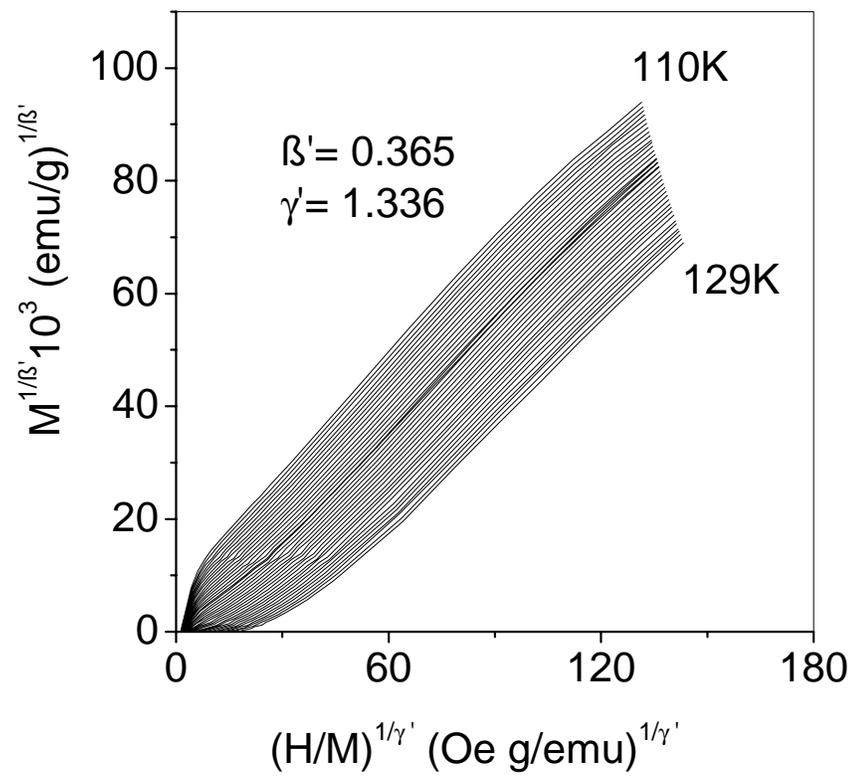

**Fig. 4**

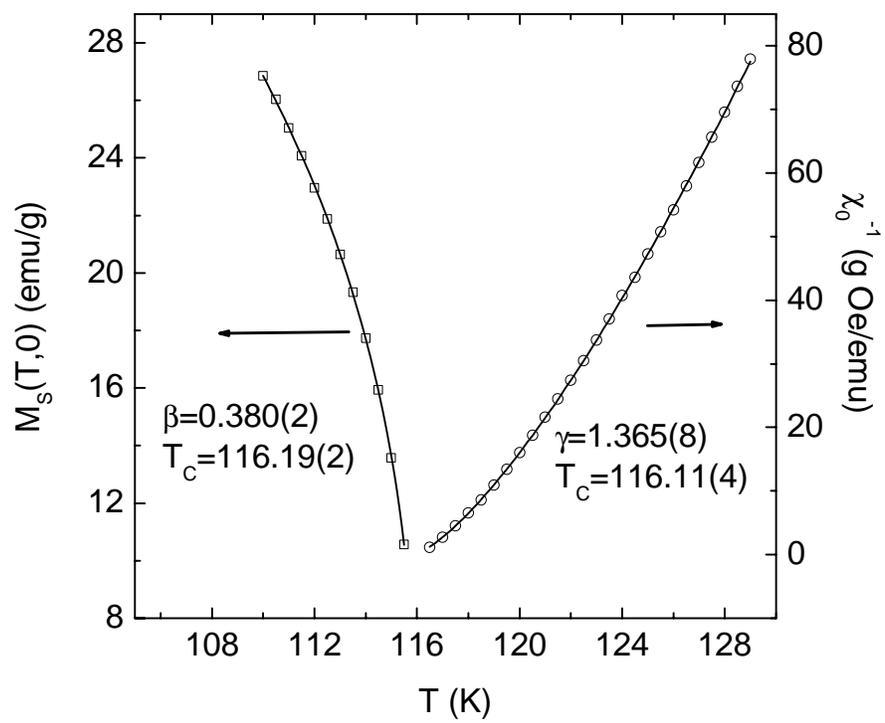

**Fig. 5**

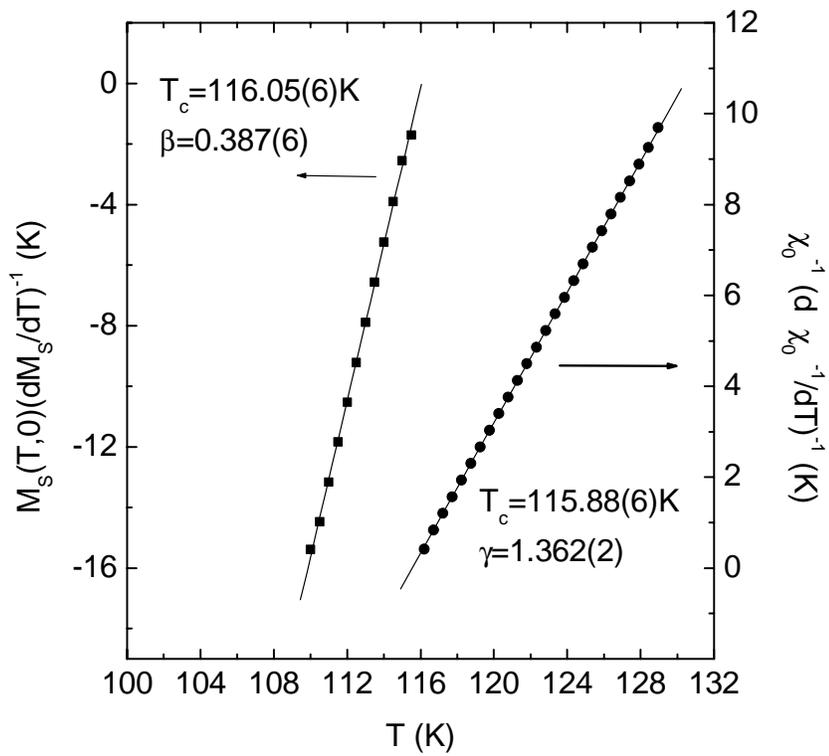

**Fig. 6**

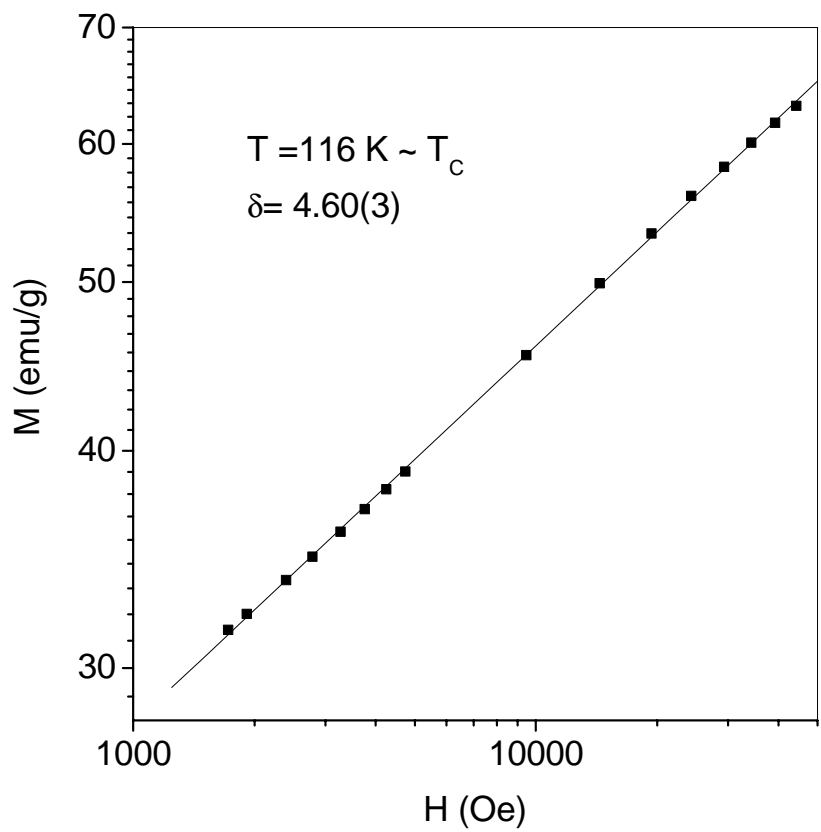

**Fig. 7**

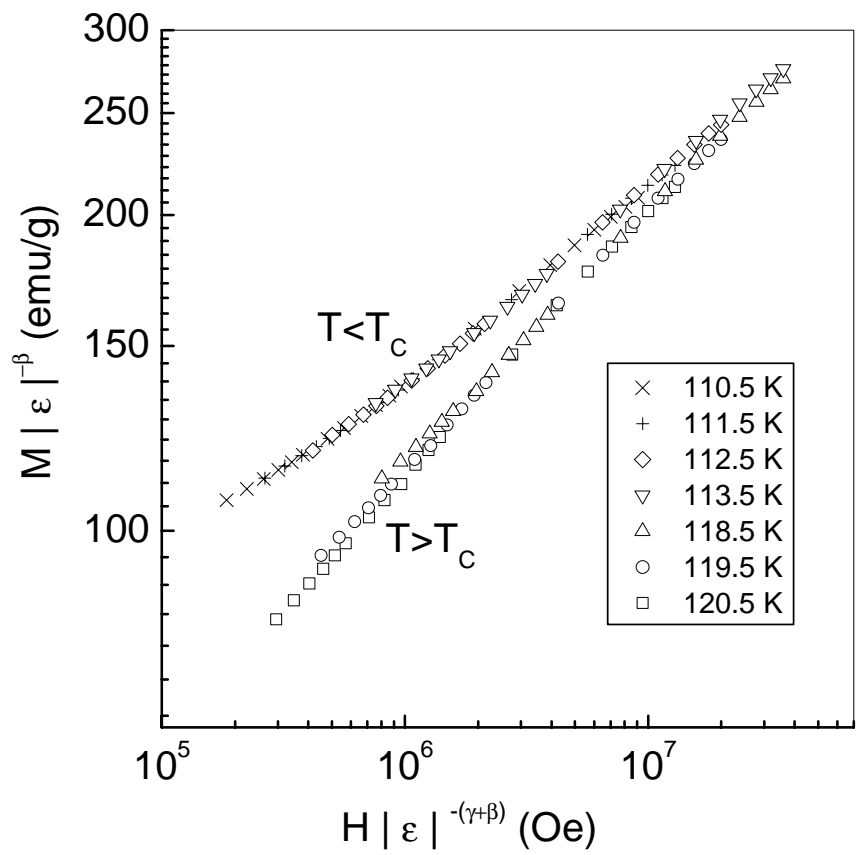

**Fig. 8**